\documentclass[sigconf]{acmart}

\usepackage{graphicx} 
\usepackage{caption}
\usepackage{subcaption}
\usepackage{xcolor}
\usepackage{balance}

\newcommand{\added}[1]{\textcolor{black}{#1}}
\newcommand{\deleted}[1]{\textcolor{lightgray}{}}

\copyrightyear{2026}
\acmYear{2026}
\setcopyright{cc}
\setcctype{by}
\acmConference[CHI '26]{Proceedings of the 2026 CHI Conference on Human Factors in Computing Systems}{April 13--17, 2026}{Barcelona, Spain}
\acmBooktitle{Proceedings of the 2026 CHI Conference on Human Factors in Computing Systems (CHI '26), April 13--17, 2026, Barcelona, Spain}
\acmPrice{}
\acmDOI{10.1145/3772318.3790527}
\acmISBN{979-8-4007-2278-3/2026/04}

\begin{document}

\title{Protocol Futuring: Speculating Second-Order Dynamics of Protocols in Sociotechnical Infrastructural Futures}

\author{Botao Amber Hu}
\orcid{0000-0002-4504-0941}
\affiliation{%
  \institution{University of Oxford}
  \city{Oxford}
  \country{UK}
  }
\email{botao.hu@cs.ox.ac.uk}

\author{Samuel Chua}
\orcid{0009-0006-0265-9572}
\affiliation{%
  \institution{Seapunk Studios}
  \city{Kuala Lumpur}
  \country{Malaysia}
  }
\email{sam@samuelchua.com}

\author{Helena Rong}\authornote{Corresponding author}
\orcid{0000-0003-1626-7968}
\affiliation{%
  \institution{New York University Shanghai}
  \city{Shanghai}
  \country{China}
  }
\email{hr2703@nyu.edu}

\begin{abstract}
Drawing on infrastructure studies in HCI and CSCW, this paper introduces Protocol Futuring, a methodological framework that extends design futuring by foregrounding protocols—rules, standards, and coordination mechanisms—as the primary material of speculative inquiry. Rather than imagining discrete future artifacts, Protocol Futuring examines how protocol rules accumulate drift, jam, and other second-order effects over long temporal horizons. We demonstrate the method through a case study of Knowledge Futurama, a multi-team participatory workshop exploring millennial-scale knowledge preservation. Using a relay format in which teams inherited and reinterpreted partially formed designs, the workshop revealed how ambiguous handovers, adversarial reinterpretations, shifting cultural norms, and crisis dynamics transform protocols as they move across communities and epochs. The case shows how Protocol Futuring makes infrastructural politics and long-run consequences analytically visible. We discuss the method’s strengths, limitations, and implications for researchers investigating emergent sociotechnical systems whose impacts unfold over extended timescales.
\end{abstract}



\begin{teaserfigure}
    \centering
    \includegraphics[width=1\linewidth]{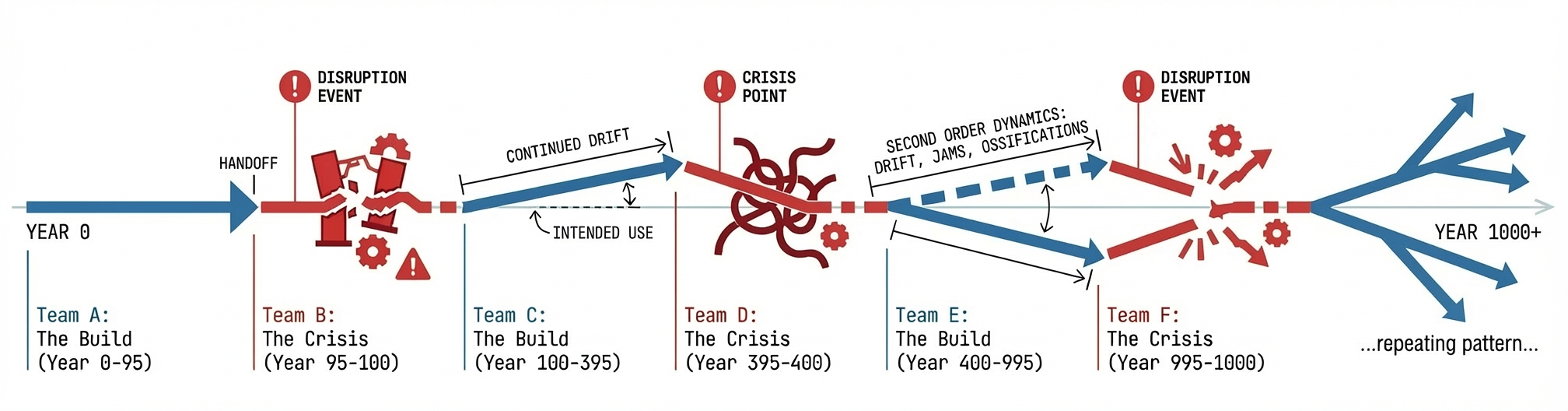}
    \caption{Protocol Futuring focuses on protocols as the primary material of speculative inquiry and operationalizes long-term infrastructural speculation through iterative handoffs between teams. In the Knowledge Futurama workshop, participants alternated between ``Blue Build'' phases (designing preservation protocols) and ``Red Crisis'' phases (introducing disruptions), passing partially-formed designs to new teams across simulated centuries. This relay structure makes visible how protocols drift, jam, and ossify as they traverse multiple hands with incomplete context—surfacing second-order dynamics that artifact-centric design futures often miss.}
    \label{fig:teaser}
    \Description{
    Protocol Futuring focuses on protocols as the primary material of speculative inquiry and operationalizes long-term infrastructural speculation through iterative handoffs between teams. In the Knowledge Futurama workshop, participants alternated between ``Blue Build'' phases (designing preservation protocols) and ``Red Crisis'' phases (introducing disruptions), passing partially-formed designs to new teams across simulated centuries. This relay structure makes visible how protocols drift, jam, and ossify as they traverse multiple hands with incomplete context—surfacing second-order dynamics that artifact-centric design futures often miss.
    }
\end{teaserfigure}

\maketitle

\section{Introduction}


In a very real sense, \emph{“the fate of our world may rest on protocols”}~\cite{Bennke2025Protological}. Technical protocols—the standards and rule sets that orchestrate reliable interactions among humans and agents—underpin planetary-scale computation and coordination~\cite{Hui2024Machine,Bratton2016stack}. They range from floating-point arithmetic used by every computing device~\cite{Lee2025Floating}, to communication protocols such as HTTP and TCP/IP, and cryptographic primitives and smart contracts like Ethereum~\cite{Zouari2025Ethereum}. Unlike products or platforms, protocols tend to \emph{“disappear”} into infrastructure~\cite{Star1996Steps}, their influence surfacing only when failures or side-effects emerge. This raises a design challenge: how can we proactively envision the futures shaped by protocols, not just the applications built atop them?

Over the last two decades, HCI researchers have widely adopted the method of \emph{“design futuring”} \cite{Celik2025Reviewing,fry2009design,Bardzell2012Critical}, such as speculative design~\cite{Auger2013Speculativea} and design fiction~\cite{Blythe2014Research} as tools to create conceptual artifacts and narratives to probe alternative sociotechnical worlds~\cite{dunne2013speculative}. Critical and speculative design \cite{Bardzell2012Critical} projects seek to probe different social values, norms, and material configurations. Yet most remain artifact-centric: they foreground near-term devices and user interactions~\cite{bleecker2009design}, while the \emph{“invisible”} infrastructures of rules and standards stay in the background. Long-term issues of evolution, maintenance, and governance thus remain implicit, prompting calls for \emph{“infrastructural speculation”}~\cite{wong2020infrastructural}: \citet{wong2020infrastructural} ask how speculative design can \emph{“explicitly center and raise questions about the infrastructures within which products are entangled”}.

Protocols \cite{summerofprotocols2025} differ from products in that their effects are usually indirect and emergent. A well-designed protocol fades into the background of a system; people interact through it, not with it. Its implications unfold through second-order dynamics in the sociotechnical system, rather than through one-to-one user interactions. For instance, when a new data-sharing protocol is widely adopted, it might initially improve interoperability (a first-order intended effect), but it can also lead to second-order phenomena: \emph{drift}, where stakeholders adapt or game the system, as encapsulated in Goodhart's law: \emph{“when a measure becomes a target, it ceases to be a good measure”} \cite{goodhart1984problems,Manheim2018Categorizing}; jams, where scaling up usage creates congestion or conflicts, echoing Frederik Pohl's observation that \emph{“a good science fiction story should be able to predict not the automobile but the traffic jam”} \cite{Pohl1973}; or \emph{ossification}, where an entrenched standard becomes resistant to change, locking in certain behaviors or power structures \cite{Ammar2018ex}. These emergent effects often only become apparent when protocols interact with complex social contexts---users, institutions, and other technologies that repurpose or stretch the rules beyond their original intent. Traditional user-centered design methods struggle to surface such long-term, systemic implications. 

To systematically explore and speculate about these hidden dynamics before they become tomorrow’s crises or lock-ins, 
in this paper, we introduce \emph{“Protocol Futuring”} (PF) as a methodological framework for critical computing and design research for infrastructure studies. Protocol futuring shifts speculative inquiry away from singular gadgets and toward the infrastructural rules and standards that coordinate sociotechnical futures. It is a framework rather than a single method: it can be realized through design fiction, participatory workshops, AI-based simulation, adversarial contests, and other research-through-design techniques \cite{Blythe2014Research}. We define Protocol Futuring through five core commitments:

\begin{itemize}
\item  Protocols over products: treat rules/standards as the primary design object.
\item  Second-order dynamics: probe drift, jams, and ossification beyond intended use.
\item  Situated lifeworlds: embed protocols in social, organizational, legal, and environmental contexts rather than abstract deployments.
\item  Multi-stakeholder argumentation: surface protocols as negotiated settlements that encode values and distribute benefits/risks.
\item  Research-through-Design and Reflexivity: think-by-making protocol probes, iteratively reflect on methods and positionality.
\end{itemize}

In operationalizing these principles, we provide a workflow that guides researchers through the process of Protocol Futuring: identifying a focal tension or infrastructural problem; constructing speculative scenarios or prototypes around that protocol to ease the tension or solve the problem; engaging stakeholders or participants to inhabit and stress-test those scenarios; and reflecting on the emergent second-order dynamics and insights gained. We detail this process in Section \ref{sec:method}, offering practical steps and considerations for applying Protocol Futuring in design research projects.

We then demonstrate the flexibility and value of Protocol Futuring through an in-depth case study, \emph{Knowledge Futurama}, a participatory adversarial workshop that explored how long-term knowledge preservation protocols evolve under crisis, reinterpretation, and drift. This flagship case allows us to “show” the method in action by tracing how teams inherited, reworked, and contested protocol designs over successive speculative centuries. To illustrate the broader adaptability of the framework, we include a supplementary case—\emph{South Beast Asia}—in the appendix. This simulated-society exercise used adjustable protocol parameters to reveal macro-social effects and unintended system-level dynamics. Across these formats, the analytic emphasis remains consistent: protocols as world-building material, the negotiation of rules across actors, and the emergence of second-order consequences that shape sociotechnical futures.

In summary, this paper contributes Protocol Futuring as a novel methodological framework for HCI, CSCW, and design research. We define the concept of Protocol Futuring and demonstrate its application across multiple scenarios, providing an infrastructural lens for speculating about sociotechnical futures. Our work offers design researchers a framework (with concrete principles and tactics) for thinking beyond individual gadgets to the rules and institutions that drive technology’s long-term impacts. By focusing on protocols and their second-order dynamics, Protocol Futuring reveals hidden politics and complex entanglements that might otherwise remain invisible in product-centered futuring. We argue that this approach enriches critical computing practice by allowing researchers to engage with questions of governance, maintenance, and systemic change in envisioned worlds. Finally, we reflect on the strengths of Protocol Futuring---such as its power to surface infrastructural power imbalances---as well as its challenges, e.g. how to validate or assess speculative insights about systems that do not yet exist.

\section{Background}

We ground the development of protocol futuring in three strands of background literature. First, we survey the emerging field of protocol studies, which examines protocols as sociotechnical phenomena and argues for their growing importance. This informs our understanding of what ``protocol'' means beyond the purely technical definition, and why protocols warrant speculative exploration. Second, we review work in speculative design, design fiction, and participatory design, tracing how these practices have evolved from focusing on artifact-centric futures towards engaging with infrastructure, policy, and plural perspectives. This situates our method among efforts to make design fictions more systemic and inclusive. Third, we draw on infrastructure studies in HCI and STS, which provide concepts that parallel our focus on the long-term life of technologies. Integrating these threads, we articulate how protocol futuring can serve as a form of critical infrastructural speculation oriented to both technical and social dimensions of future networks.

\subsection{Protocols and Protocol Studies}

\subsubsection{Technical Protocol Design Practices}

Technical protocols are the backbone of interoperable technology, prescribing how systems communicate and function. Classic Internet design principles exemplify this culture of protocol development. The Internet Engineering Task Force (IETF) famously embraces \emph{“rough consensus and running code,”} a maxim that privileges practical implementations and community agreement over formal authority~\cite{Russell2006Rough}. David Clark’s dictum—\emph{“We reject: kings, presidents, and voting. We believe in: rough consensus and running code”}—encapsulates the preference for open, collaborative iteration in protocol design. Protocols such as TCP/IP were deliberately minimal, \emph{“caring nothing for what is said or who said it,”} to remain flexible building blocks~\cite{galloway2004protocol}. These values of simplicity, modularity, and extensibility have been credited with the Internet’s success, even as they separated social or policy considerations from early engineering work~\cite{denardis2014global}.


\subsubsection{Protocols as Sociotechnical Infrastructures}

Computing definitions treat protocols as rules for data transmission, but interdisciplinary scholarship broadens the concept to sociotechnical coordination mechanisms—\emph{“codified behaviors”} enabling complex, distributed activity~\cite{galloway2004protocol}. The Summer of Protocols initiative\footnote{\url{https://summerofprotocols.com/}} adopted and debated Danny Ryan’s definition of a protocol as \emph{“a stratum of codified behavior that allows for the construction or emergence of complex coordinated behaviors at adjacent loci,”} eventually condensing it to the pithy insight that \emph{“a protocol is an engineered argument”}~\cite{summerofprotocols2025}. In this view, every protocol encodes a negotiation of values and priorities, shaping collaboration and conflict alike.

Protocols often fade into the background when functioning well, a phenomenon Alfred North Whitehead anticipated when he wrote that \emph{“civilization advances by extending the number of important operations we can perform without thinking of them”}~\cite{Whitehead1911}. They coordinate actions at scales ranging from the mundane (traffic lights, handshakes) to the global (TCP/IP, blockchain consensus), yet typically draw attention only when they fail~\cite{star1999ethnography}. Scholars stress that protocols are infrastructural forces on par with markets or organizations but remain undertheorized and under-taught~\cite{pipek2009infrastructuring}. Engineers may not learn to read or write protocol specifications formally, and most social-science frameworks do not treat protocols as first-class analytical objects.

Despite their hiddenness, protocols carry social and political weight. \citet{lessig1999code}’s dictum that \emph{“code is law”} captures how software rules can regulate behavior as effectively as legal statutes. Contemporary analyses extend this logic to \emph{“social protocols”}—the explicit and tacit rules that structure interaction online and off~\cite{warden2023social}. Platform algorithms, moderation policies, and community norms jointly govern information flows, demonstrating how technical and social layers of protocol interlock. \citet{masnick2019protocols} argues that shifting from proprietary platforms to open protocols could address persistent challenges in content moderation and free speech online, proposing that protocol-based architectures give users more choice and create healthier competitive dynamics than centralized platform governance. Formal standard-making bodies increasingly treat social values as matters of codification. ISO 14001 (``Environmental management systems'') and ISO 26000 (``Social responsibility''), for example, establish environmental and social responsibility protocols that go beyond conventional law~\cite{Yates2019}.

Yet the very success of protocols breeds neglect. Effective protocols recede into infrastructure, making failures hard to anticipate and reforms difficult to introduce. Internet history offers cautionary examples of \emph{“protocol ossification,”} \cite{Ammar2018ex} where widely deployed standards become so entrenched that even beneficial upgrades face rejection by middleboxes or firewalls~\cite{handley2015ossification}. More generally, neglected protocols can grow fragile, sclerotic, or vulnerable to capture by narrow interests or malicious actors. \citet{askonas2025control} describes how protocols create \emph{``accountability sinks''}---situations where decisions are delegated to policies or automated systems such that no human appears responsible or in charge. In a protocol-governed society, agency becomes invisible: the more decentralized and scalable the system, the more responsibility evaporates, even as power continues to operate~\cite{masnick2019protocols}. Conversely, well-designed and well-stewarded protocols can enable major leaps in coordination, as rail gauges, electrical standards, and Internet protocols did in earlier eras.

\subsubsection{Standardization Processes: History and Institutional Context}
Protocols and standards emerge through institutionalized negotiations that balance innovation, consensus, and power. \citet{Yates2019} trace three historical ``waves'' of global standard setting. In the late 19th and early 20th centuries, professional societies such as ASTM and IEEE developed technical specifications to enable industrial interoperability. Mid-20th-century efforts like the ISO and ITU expanded coordination globally. From the late 1990s onward, a third wave has addressed broad governance issues—human rights, environmental stewardship, social responsibility—treating values themselves as subjects of standardization.

Different arenas exemplify different governance models. The IETF relies on an open, volunteer-driven RFC process, while ISO and ITU operate formal, consensus-based negotiations among governments and corporations. Each approach trades off agility and political legitimacy. The enduring challenge, as \citet{hanseth1996tension} noted, is balancing stability and flexibility so that standards remain universal yet adaptable. Rigid standards can become barriers to change, while excessive flexibility undermines interoperability.

Standards can also spur creativity rather than stifle it. \citet{jackson2015standards}’s study of an ecology research network showed that defining shared data and workflow protocols was a generative design activity, coining the idea of \emph{“standards and/as innovation”}. By bringing diverse actors together to formalize solutions, standardization can itself be a site of discovery and collaboration.

\subsubsection{Speculative and Fictional Protocols}

Recently, \emph{Protocol fiction} has emerged as a distinct approach that foregrounds protocols themselves as the primary object of speculation. \citet{webb2023protocol} defines protocol fiction as a method for imagining new infrastructure---such as drone delivery networks or health screening ecosystems---by specifying protocols that enable cooperation among aligned actors and articulating both the \emph{belief} (a plausible path to implementation) and \emph{desire} (a compelling visualization of the future) needed for adoption. Unlike conventional design fiction, which often centers on discrete artifacts, protocol fiction treats the coordination mechanism itself as the speculative material.

This orientation toward protocols as narrative material has gained traction through initiatives like the \emph{Terminological Twists} fiction contest, which challenged writers to take existing protocol-related technical concepts and reverse-engineer them into compelling science fiction stories~\cite{protocolized2025terminological}. The contest, which attracted fifty submissions, exemplifies protocol fiction as \emph{``a fresh subgenre of sci-fi, with systems and rules almost as protagonists in their own right''}~\cite{protocolized2025guidelines}. Guided by what practitioners call ``Chiang's Law''---that science fiction is about \emph{strange rules} while fantasy is about \emph{special people}---protocol fiction explores regimes of rules and the dynamics of worlds that grow on those protocols. Winning entries such as \emph{``Noise Ordinance,''} \emph{``The 40-Hour Work Week,''} and \emph{``DHCP''} demonstrate how technical protocol concepts can anchor speculative narratives that probe social and political implications.

The Summer of Protocols initiative employs protocol fiction as a means of provocation and exploration. For example, \emph{Composable Life} envisions blockchain as an \emph{``unstoppable nature''} that gives rise to artificial life, leveraging speculative narrative and immersive virtual reality to probe the long-term evolution of protocols and the intricate, symbiotic relationships between humans, AI, and blockchain technology~\cite{hu2025composable}. In a similar vein, \emph{Autonomous Realities} introduces speculative protocols for shared mixed reality experiences, including an embodied protocol for establishing MR sessions  inspired by ancient secret handshakes, alongside a meta-protocol for sustaining persistent, autonomous shared reality layers~\cite{hu2025autonomous}. These projects push protocol fiction beyond traditional narrative, extending it into the realms of experiential and embodied interaction.

HCI research has actively engaged with speculative approaches to protocols and infrastructure. For instance, \citet{rodden2013smart} deployed “agents” within homes to simulate future smart energy grids, probing how people might interact with and perceive emerging energy infrastructures. \citet{soden2019infrastructuring} explored how communities practice \emph{“infrastructuring the imaginary”} in response to challenges like sea-level rise, highlighting the imaginative work underpinning collective adaptation. Similarly, \citet{jabbar2019blockchain} demonstrated that introducing blockchain technologies compels organizations to reimagine workflows and reshape their infrastructural imaginaries. Some projects take an inverse approach, using the absence of protocols as a critical provocation: \citet{grandhi2020internetless} posed the question \emph{“An Internet-less World?”} to prompt reflection on society’s profound dependency on network protocols, illustrating how scenarios of protocol failure can serve as productive sites for creative inquiry.

These experiments treat protocols not only as technical artifacts but also as sites for critical imagination. By dramatizing potential breakdowns, alternative coordination mechanisms, or speculative futures, they surface hidden assumptions, ethical dilemmas, and opportunities for resilience that conventional planning might miss---providing a foundation for our own methodological contribution.

\subsection{Design Futuring}

Speculative design and design fiction have emerged over the past two decades as important methods in HCI and design research for interrogating future possibilities \cite{Celik2025Reviewing}. These approaches create fictional artifacts, scenarios, or narratives that encourage audiences to question assumptions and consider alternative social or technological trajectories. As originally articulated by designers like ~\citet{dunne2013speculative}, speculative design emphasizes provoking critical reflection on social values and ethics by presenting uncanny or exaggerated design concepts (often as physical prototypes or visual media). Design fiction, a term popularized by science fiction author Bruce Sterling and designer Julian Bleecker, is closely related: it involves using fictional diegetic prototypes---artifacts that exist within an imagined future world---to \emph{``suspend disbelief about change''} and spark discussion~\cite{sterling2005shaping,bleecker2009design}. Sterling notably described design fiction as \emph{``the deliberate use of diegetic prototypes to suspend disbelief about change''}, suggesting that its power lies in focusing people on a specific object or service in context, rather than on abstract futurism or grand policy. In practice, design fictions often take the form of videos, stories, or concept designs that ask \emph{``what if...?''} about near-future technologies---for example, envisioning new gadgets and their everyday use, or hypothetical tech policies and their societal impact. A recent systematic literature review identifies four core phases common to speculative design processes: select, explore, transform, and provoke~\cite{Cardenas2025Systematic}.

\subsubsection{Reflective Modes and Critical Perspectives}

\citet{Kozubaev2020Expandingb} contribute five reflective modes to help HCI researchers improve design futuring work: designerly form-giving (the specificity and experiential qualities of design objects), temporality (how futures are represented across time), positionality (recognizing that futuring emerges from particular situated perspectives), real-world engagement (clarifying how speculation connects to actual contexts), and knowledge production (understanding what kinds of knowledge design futuring generates). These modes function as analytical resources for practitioners to articulate, evaluate, and generate new design futuring work, addressing methodological challenges as design futuring becomes more established in HCI.

\citet{Howell2021Plurality} critique the Futures Cone---a prominent model in design futuring---for reducing diverse perspectives to a singular point of ``the present'' and embedding assumptions about linear progress. This work responds to calls for decolonizing futures approaches by acknowledging that there is no single everyday experience and that assuming universality obscures the complexity of historical context and diverse, situated presents. 

A recent literature review of HCI's future-orientation~\cite{Sanchez2025LetsTalk} argues that while HCI is inherently future-oriented, the visions of futures in HCI publications tend to be implicit, techno-deterministic, narrow, and lacking in roadmaps and attention to uncertainties. However, a growing body of HCI research is challenging techno-centric visions by demonstrating active exploration of uncertainty, focus on human experience, and contestation of dominant narratives. \citet{Bendor2024Teaching} address speculative design as pedagogical practice, examining how to integrate this ``critical material engagement with possible futures'' into design-engineering curricula and identifying obstacles for curriculum integration.

\subsubsection{Participatory Speculative Design}

Participatory design (PD), on the other hand, originates from a different tradition and focuses on involving stakeholders---especially end users or marginalized groups---directly in the design process. The core idea of PD is to democratize design decision-making and ensure outcomes reflect the needs and values of those affected~\cite{simonsen2012routledge}. While classic PD was applied to developing real systems, there has been growing interest in participatory speculative design: inviting participants to co-create visions of the future. This fusion aims to overcome one criticism of early speculative design and design fiction: that they sometimes reflected the narrow, ``privileged'' perspectives of their authors~\cite{tonkinwise2014design}. Scholars have pointed out that many speculative design works, especially from Western academic or art contexts, carried assumptions not shared by broader populations, and risked a kind of cultural or technological elitism. In response, methods like participatory design fiction (e.g. workshops where diverse groups build fictional scenarios together) and speculative co-design have been developed to inject plurality and ground speculation in a wider range of experiences~\cite{lindley2014anticipatory,gerber2018participatory}. 

A systematic literature review on participatory speculative design identifies this practice as shifting focus from artifacts to process, empowering public participation across multiple stakeholders to ``co-create potential pluralistic futures and democratize imagination''~\cite{Zhang2024Pluralistic}. This research outlines three practical pathways---technical speculation, social speculation, and integrated speculation---along with seven participatory methods and a four-phase framework to guide participatory speculative design practice. 

\subsubsection{From Artifacts to Infrastructures}

Another notable evolution in design fiction practice has been a shift from focusing on single artifacts to considering systems and infrastructures. Sterling's notion of keeping design fiction \emph{``on the level of objects and services''} was a useful starting heuristic, but many contemporary design researchers have sought to reconnect those isolated artifacts to the larger contexts that give them meaning. As \citet{wong2020infrastructural} observe, speculative designs \emph{``often imply a lifeworld''} even if they only show a gadget. There is an increasing push to explicate that lifeworld: to design not just a speculative object but also illustrate the networks, practices, regulations, and cultures around it. Infrastructural Speculations \cite{wong2020infrastructural} exemplify this trend. They outline design tactics to deliberately incorporate infrastructure considerations into speculative design, so that the speculation \emph{``places lifeworlds at the center of design concern''} and highlights the cultural, regulatory, environmental, and repair conditions enabling the imagined future. Their work suggests techniques like \emph{``infrastructural inversion''}---essentially flipping the focus to the normally invisible support systems---and considering multiple stakeholders and temporal scales in speculative scenarios.

\citet{Wakkary2015Material} propose material speculation as a complementary approach that ``utilizes actual and situated design artifacts in the everyday as a site of critical inquiry.'' Rather than relying solely on scenarios, prototypes, and forecasting, material speculation emphasizes using real, tangible artifacts to provoke critical reflection by coupling counterfactual artifacts---objects designed to challenge everyday logic---with literary concepts of possible worlds. 

\subsection{Infrastructure Studies, Temporality, and Sociotechnical Imaginaries}

Our work is additionally informed by the rich body of infrastructure studies in both HCI and Science \& Technology Studies (STS). Infrastructure in this context refers to the large-scale sociotechnical systems and underlying structures that support human activities---from transportation grids and power networks to information systems and organizational routines. A foundational perspective by \citet{Star1996Steps} reconceptualized infrastructure not as a static thing, but as something that is relational, ecological, and learned through use. \citet{bowker2010toward} argue for \emph{``information infrastructure studies''} as a way of attending to the mundane yet consequential work of standardization, classification, and data curation that underpins contemporary knowledge practices. \citet{edwards2003infrastructure} outlines an agenda for infrastructure studies that emphasizes heterogeneity, embeddedness, and the long temporal horizons of infrastructural systems. \citet{Ribes2007Tensions} describe the long-term planning of infrastructure development, highlighting tensions between innovation and stability, local flexibility and global interoperability, short-term projects and long-term maintenance.

A recent systematic literature review by \citet{lyu2025infrastructure} identifies three major themes in SIGCHI infrastructure research:
\begin{itemize}
    \item \textbf{Growing infrastructure}: How infrastructures are developed, expanded, maintained, and repaired over time. This includes initiation of new infrastructures, sustaining them, and the collaborative work (by engineers, users, maintainers) that keeps them going. For example, studies of open-source communities or civic tech often look at how volunteer contributors build and sustain a technical infrastructure over years.
    \item \textbf{Appropriating infrastructure}: How people use and adapt infrastructures in practice, often in ways not foreseen by the designers. This covers the inventive repurposing of systems and the integration of formal infrastructure with informal practices to facilitate collaboration and participation. For instance, citizens using social media platforms (built as commercial infrastructure) for crisis response and mutual aid would be an appropriation phenomenon.
    \item \textbf{Coping with infrastructure}: How people respond when infrastructures break down, constrain, or fail to support them. Such studies examine the workarounds, improvisations, and alternative solutions people devise to overcome infrastructure shortcomings or outages. A classic example is how communities handle electrical blackouts or how patients and clinicians cope when a healthcare IT system goes offline---effectively creating ad-hoc infrastructures to bridge the gap.
\end{itemize}

Subsequent work has elaborated concepts such as infrastructure time and infrastructuring. \citet{karasti2004infrastructuring} examine how collaborative data infrastructures in science require ongoing care, negotiation, and reconfiguration over years and decades, rather than one-off design interventions. Infrastructuring, drawn from participatory design, foregrounds users and communities as co-designers who continuously shape, appropriate, and re-align infrastructures in practice \cite{karasti2004infrastructuring,bjorgvinsson2012agonistic}. \citet{Larkin2013Politics} analyzes \emph{``the politics and poetics of infrastructure''}, showing how infrastructures materialize state power, cultural imaginaries, and sensory experiences. Anthropological accounts of infrastructure in cities, energy systems, and logistics further emphasize uneven access, differential vulnerabilities, and the improvisational practices people develop to cope with breakdowns \cite{anand2018promise,Harvey2012Topological}.

Crucially for our purposes, STS has also foregrounded temporality and sociotechnical imaginaries as central to infrastructural politics. \citet{jasanoff2015dreamscapes} conceptualize sociotechnical imaginaries as collectively held, institutionally stabilized visions of desirable futures that guide large-scale techno-political projects. \citet{sareen2021matter} argue that studying energy infrastructures requires explication of temporality, including how expectations, delays, and irreversibilities shape what futures become possible. Infrastructure studies thus provide both conceptual tools (e.g., inversion, infrastructuring) and temporal sensibilities (e.g., long-now, path-dependence, legacy) that are highly relevant to protocol design.

\section{Methodological Framework: Protocol Futuring}
\label{sec:method}

Protocol Futuring is a methodological framework for speculating on the future of sociotechnical infrastructures by using design fiction techniques centered on protocols. In essence, it asks researchers, designers, and participants to imagine future worlds where the key drivers or points of inquiry are protocols---whether technical standards, social rules, or political processes---and to explore the second-order consequences of those protocols in action. This approach extends speculative design into what we call an ``infrastructural lifeworld'' mode, where the unit of analysis isn't a single artifact or user experience but rather the network of relationships, agreements, and hidden work that a protocol coordinates or disrupts. This situates our method within, but distinct from, established speculative design traditions \cite{Star1996Steps,bowker1999sorting,pipek2009infrastructuring}.

There are several defining objectives and characteristics of Protocol Futuring:

\paragraph{Centering "Strange New Rules"} In contrast to previous work, our primary aim is to shift speculative design's focus from products to protocols. We concentrate on new protocols and how they function in future lifeworlds. Rather than emphasizing specific individuals or heroes in our speculative fiction, we focus on the systemic effects of protocols. Our approach is guided by what we call Chiang's Law, named after science fiction author Ted Chiang \cite{Rao2025Strange}: \emph{``science fiction is about strange rules, while fantasy is about special people.''} This aligns the method with recent moves in HCI to examine governance-by-infrastructure, standards, and rule systems \cite{lampland2009standards, ribes2017notes}, yet uses fiction to expose how those rules behave over extended temporal horizons.
    
\paragraph{Foregrounding Second-Order Effects} A primary aim is to move beyond first-order effects (``new technology X performs function Y'') to examine second-order and longer-term effects (as X becomes widespread, it causes Z unintended outcome among society/environment, or actors respond with W adaptation). Second-order effects are explored not as deterministic predictions but as plausible dynamics grounded in what \citet{edwards2003infrastructure} calls the "vast machine" of infrastructural interdependence. The method deliberately directs attention to what happens after a protocol's initial deployment and uptake—essentially, to consider the trajectory of a protocol through time (e.g., years, decades) and scale (from small pilot to global infrastructure). This exploration includes phenomena like protocol drift (implementations or uses diverging from original design over time), protocol jams (congestions or deadlocks in socio-technical systems caused by interacting protocols, similar to traffic jams), and protocol ossification (protocols becoming rigid and resistant to change as they entrench). For example, a Protocol Futuring about a future internet standard might explore how that standard initially improves security but later leads to a new form of digital lock-in or black market to circumvent its restrictions—outcomes beyond the protocol's immediate goal. By targeting these ripple effects, Protocol Futuring aligns with the ethos of \emph{``predicting the traffic jam, not just the car''}~\cite{Pohl1973}, using imaginative narrative as a foresight tool. It is worth noting that here, we treat drift, jam, and ossification not as predictions or diagnoses, but as interpretive hypotheses generated through speculative enactment. These dynamics function as candidate patterns that invite later empirical investigation rather than claims about inevitable outcomes.
    
\begin{quote}
\emph{“A good science fiction story should be able to predict not the automobile but the traffic jam.”} – Frederik Pohl
\end{quote}

\paragraph{Lifeworld and Context at Center} 
In Protocol Futuring, context is not backdrop but a primary analytic unit: scenarios are constructed to show how a protocol operates within—and is reshaped by—the social norms, institutions, regulatory conditions, and legacy infrastructures that constitute its lifeworld. This aligns the method with infrastructural speculation \cite{wong2020infrastructural} and with feminist and care-centered infrastructuring traditions that foreground ongoing maintenance and situated negotiation \cite{pipek2009infrastructuring,karasti2014infrastructuring}. Rather than presenting protocols as neutral technical rules, Protocol Futuring depicts how they are interpreted, contested, and maintained over time. This may involve narrative techniques such as following invisible actors (e.g., maintainers, governance committees) or describing scenes from the protocol’s point of view. The aim is to make the infrastructural “wires and threads” visible and to situate protocols within the lived practices that give them force.

    
\paragraph{Multi-Stakeholder and Participatory Approach} 
Because protocols distribute authority across designers, implementers, users, regulators, adversaries, and bystanders, Protocol Futuring relies on participatory methods to surface these divergent viewpoints. Building on traditions in participatory design \cite{ehn2008participation,bjogvinsson2012design} and game-based co-design \cite{gaver1999design,brandt2012tools}, our approach invites multiple contributors to co-construct speculative scenarios through workshops, role-play sessions, and open fiction calls. This distributes authorship beyond designers and enables participants to inhabit different stakeholder positions---everyday users, maintainers, marginalized groups, or adversaries---thereby generating plural, sometimes conflicting, worldviews. Such plurality aligns with the need to imagine protocols that require collective adoption and negotiation rather than heroic individual use. In practice, we draw on PD facilitation techniques (e.g., scenario mapping) and design-game mechanics \cite{brandt2012tools} to structure interaction. These participatory and game-like structures make the infrastructural politics of protocols experientially accessible and foreground how rules shape—and are shaped by—collective imagination.

\paragraph{Artifacts and Media} Although protocols can seem abstract, Protocol Futuring materializes speculative worlds through concrete artifacts, \added{drawing from traditions in critical design \cite{dunne2013speculative}, design fiction \cite{sterling2005shaping}, and future-oriented prototyping \cite{halse2010rehearsing}.} Outputs may include short stories, fictional policy documents, speculative diagrams, interface mockups, or immersive videos—forms that make the protocol’s logic, breakdowns, and sociopolitical effects legible. These artifacts function not as ``hero objects'' but as probes that reveal how a rule system organizes everyday life. For instance, a fabricated citizen handbook for a future data protocol or a visualization of rerouted network flows during a protocol split can foreground operational constraints, power dynamics, and maintenance work. We often employ cross-media storytelling—text, images, interactive maps, AI-generated video—to surface different infrastructural layers. As demonstrated in the South Beast Asia case presented in the Appendix, such multimodal artifacts help participants inhabit the protocol’s world and expose second-order dynamics that would remain invisible in purely textual speculation. Protocol Futuring thus describes not a fixed output format but a strategic orientation capable of using any medium necessary to render protocol-driven futures thinkable.

    
\paragraph{Research-through-Design and Reflexivity} Methodologically, we treat Protocol Futuring as a form of research-through-design (RtD) \cite{Blythe2014Research}. This means the act of creating the fiction is itself a mode of inquiry: by designing these speculative scenarios, we are effectively probing research questions about protocols (``what if this protocol were in place, how would X or Y play out?'') and generating knowledge in the form of narrative insights, identified pitfalls, or design considerations, following \citet{zimmerman2007research}. Throughout our process, we maintain a reflexive stance---we document the decision-making, the alternatives considered, the influences from real-world references, and the participants' reactions. Analysis of a protocol fiction may involve looking for emergent themes across different stories, or using the fiction as a provocation in interviews with experts. For example, after producing a set of stories in the Knowledge Futurama workshop, we conducted a debrief discussion where librarians and archivists (some of whom participated) reflected on the scenarios: which seemed plausible, which touched on real concerns, which surprised them by revealing a blind spot in current practices. This reflective step, \added{aligning with \citet{ghajargar2021synthesis},} ensures that the fictional outputs are brought back to bear on present understanding, closing the loop of speculation to inform reality.

\subsection{Process Overview}

In practical terms, a typical Protocol Futuring engagement follows six steps:

\begin{enumerate}
    \item \textbf{Choose a Focal Tension}: The process begins by identifying a sociotechnical tension that is both technically plausible and socially consequential. This may be an emerging real-world challenge (e.g., a next-generation social media standard or climate treaty framework) or a hypothetical scenario addressing competing stakeholder interests. Effective focal tensions involve trade-offs that cannot be resolved through technical design alone---such as authenticity versus interpretability, liability versus autonomy, or privacy versus auditability. The framing can be specific or broad (e.g., ``imagine the future of education protocols'') depending on the desired scope. The protocol, conceived as an engineered argument, then becomes a vehicle for navigating these tensions.
    
    \item \textbf{Gather Research and Inspiration}: Before engaging in fiction, participants compile background research on relevant technology trends, historical precedents, and social contexts. For example, speculation about knowledge archive protocols might draw on current digital preservation efforts, historical library systems, and cultural narratives about archives. This grounding anchors the fiction in reality, ensuring that speculative extrapolations remain relatable and plausible.
    
    \item \textbf{Workshop or Creative Session}: Participatory engagements typically involve a structured session---often half a day---that brings together stakeholders and creative practitioners. Facilitation exercises guide participants through stakeholder mapping, tension identification (e.g., privacy versus transparency, centralization versus decentralization), and future brainstorming. A common technique is the scenario matrix: participants vary two key factors to generate four quadrant futures, then develop narratives within each quadrant.
    
    \item \textbf{Protocol Creation}: Participants develop fictional protocols by articulating their rules and embedding them within narrative contexts. We encourage depicting concrete lifeworld situations---a day in the life of someone governed by the protocol, a news report about a protocol failure, or a diary entry from a system maintainer. These situated narratives enable imaginative exploration within the constraints established by the protocol rules. Artifacts may include explanatory annotations as needed.
    
    \item \textbf{Speculation and Analysis}: Following creation, participants analyze the fictions to surface second-order dynamics. This analysis examines whether independent narratives converge on particular risks, or whether they propose design features worth further exploration (e.g., kill-switches, consent mechanisms, fallback modes). In participatory settings, this analytical work occurs in real time as participants inherit and respond to prior teams' designs---interpreting ambiguous handovers, inferring unstated rationales, and deciding whether to extend or contest emerging protocol worlds.
    
    \item \textbf{Iterate and Refine}: Initial rounds often reveal blind spots, such as scenarios that are overly optimistic, excessively technical, or insufficiently attentive to social conflict. To address this, we introduce adversarial iteration: a ``Blue Team'' extends or stabilizes the existing protocol world, while a ``Red Team'' challenges assumptions, surfaces vulnerabilities, and explores adversarial interpretations. Because fiction accommodates multiple coexisting truths, iteration does not require converging on a single ``correct'' future. Instead, the design space is refined through counter-fictions, competing perspectives, and revisions that expose alternative trajectories.
\end{enumerate}

Through these steps, Protocol Futuring serves as a sandbox to experiment with the future of coordination. It allows us to simulate---in a low-risk narrative space---how complex socio-technical interventions might play out. In doing so, it can surface ethical questions and sociopolitical dynamics early. For example, one might discover through fiction that a theoretically decentralized protocol could paradoxically create new central points of control (like certificate authorities or moderators) which then become power bottlenecks---a realization that can inform real protocol design to mitigate that issue. Or one might explore the long tail of maintenance: who patches the system in 20 years when original developers are gone? Fiction can personify that future maintainer and give them voice now.
\section{Case Study: Knowledge Futurama}

To demonstrate the utility and versatility of Protocol Futuring, we focus in depth on a flagship case: Knowledge Futurama, a multi-team participatory fiction exercise that most clearly embodies the framework introduced in the previous section. Knowledge Futurama allows us to trace, with analytic detail, how protocol rules, mediating practices, and second-order effects emerge through iterative inheritance and reinterpretation—making it an ideal primary case for illustrating the method’s analytical contribution. Knowledge Futurama centers on long-term knowledge preservation and highlights the multi-century drift of infrastructural protocols. 

\subsection{Context and Goal of Knowledge Futurama}

Knowledge Futurama was a participatory futures workshop conducted as part of the Summer of Protocols 2025 \footnote{\url{https://summerofprotocols.com/}} program at the Edge Esmeralda \footnote{\url{https://www.edgeesmeralda.com/}} gathering in California. Around thirty participants—including technologists, designers, librarians, artists, and futurists—explored how knowledge might be preserved and made meaningful across a millennium. The aim was to prototype \emph{“protocols for 1,000-year libraries”}: sociotechnical arrangements that could keep knowledge alive, intelligible, and legitimate through centuries of political, ecological, and technological change. The name echoed the 1939 World’s Fair “Futurama” exhibit, but turned its speculative lens toward long-term memory infrastructures rather than consumer futures. 

Two of the authors participated directly in the workshop as embedded researchers. We adopted a participant-observer stance \cite{dourish2006implications}, documenting the evolving scenarios, mediating practices, and interpretive ambiguities as they unfolded. Following the event, we conducted elicitation interviews with two workshop participants (one of whom was an author of this article, indexed as “Interviewee 2”) to reconstruct design rationales, clarify interpretive moves, and surface participants’ own theories of preservation, drift, and relevance. These interviews were not intended to provide a representative sampling but to generate situated, experiential insight into how actors reasoned through long-term protocol design and how the relay structure shaped their thinking. We framed the core design challenge as building a preservation system that is not simply about storing data, but about maintaining usable protocols that can evolve. Participants emphasized a nearer design horizon of 100 years as a meaningful unit for institutional and infrastructural commitments, while also using speculative devices to explore what happens as protocols outlive their creators.

\subsection{Process and Setting}

The workshop was structured as a relay across six parallel story worlds. Participants were divided into tables of five and worked through six sequential rounds, with each table contributing to every scenario in turn. After the sixth round, each story was returned to its original authors. This allowed them to see how their initial protocol had drifted, mutated, or been reinterpreted by others, rendering interpretive drift visible and allowing them to experience how protocols evolve when traversed across many hands with only partial context.

Each story world began with a “Blue Build 1” phase, where a team proposed a knowledge-preservation protocol for the next 100 years. This required teams to articulate both a technical stack (e.g., encoding methods, storage substrates, verification technologies) and a social milieu (e.g., maintenance institutions, stewarding communities, norms, governance practices). In the subsequent round, a new team inherited this design and introduced a “Red Crisis 1” intervention that represented political, ecological, social, or technological shocks that destabilized or reinterpreted the original protocol. The third round, Blue Build 2, tasked yet another team with continuing the story under these altered conditions for the next 300 years. They had to adapt, improvise, or recompose the preservation protocol in light of the disruptions and ambiguities left by their predecessors. This was followed by Red Crisis 2, in which yet another group introduced further disruptions, forcing reinterpretations, repair, or workaround practices. This alternating pattern of “build” and “crisis” repeats one last time with Blue Build 3 extending the protocol into the far future (the next 600 years), after which Red Crisis 3 introduced a last set of shocks, before the story was returned to the original authors. Over the course of this process, each knowledge artifact developed a layered, multi-century “protocol biography”: a record of how initial rules, assumptions, and incentives accumulated unintended consequences over centuries.

The workshop was conducted entirely through six shared Google Doc files, one per story world, edited sequentially by all thirty participants. Each table deliberated in person while one designated note-taker captured its decisions around technology stacks, social milieus, crises, and state conditions directly into the evolving document. Because teams worked from what previous groups had written – brief bullet points, shorthand, half-formed sentences, and ambiguous phrasing – later teams frequently had to infer missing rationale, decide whether a line was meant as a firm rule or a tentative thought, and interpret text that reflected the stylistic quirks of an earlier note-taker rather than any stable consensus. No communication was allowed between tables until the final debrief, amplifying the interpretive gaps and making drift a constitutive part of the design process.

This structure deliberately reproduced the dynamics of protocol evolution in the wild, where specifications, proposals, patches, and lived practices often sediment into infrastructures without any single authorial center or developmental trajectory. As in actual decentralized systems, protocols themselves were the design artifacts, rather than the fictional worlds, but the rules, mediations, and assumptions that structured their long-term behavior. The shared documents functioned as proto-infrastructures that were mutable, cumulative, and open for reinterpretation, reflecting how technical rules and social practices co-produce one another over time.

Participants described this experience in the interviews as “working from someone else’s half-finished protocol” (Interviewees 1 and 2). The relay mechanism, ambiguous inheritance, and speculative time horizons made protocol drift palpable: teams saw how their design intentions were reinterpreted, contested, or overwritten by actors with different assumptions and incomplete information. This is the analytic signature of Protocol Futuring: protocols are not abstract “rules,” but dynamic socio-technical ensembles that evolve as they circulate through hands, accrue meaning, and sediment into the worlds they help create.

Interpretation of the fictional outputs occurred at multiple stages. During the workshop, participants themselves engaged in real-time interpretation as they inherited and responded to prior teams' designs—making sense of ambiguous handovers, inferring unstated rationales, and deciding how to extend or contest the emerging protocol worlds. In the post-workshop phase, the research team conducted follow-up interviews and collaborative analysis sessions with participants, during which both researchers and participants reflected on the scenarios, surfaced implicit assumptions, and identified recurring patterns such as drift, jam, and ossification. This layered interpretive process—participant sense-making during the relay, followed by researcher-participant collaborative analysis—grounds the methodological insights in both experiential and analytic registers.

\subsection{Illustrative Story: Wikipedia and the “WikiComet”}

To illustrate how Protocol Futuring operates in practice, we focus on one of the story worlds in which both our interviewees participated: a speculative effort to preserve Wikipedia by launching a millennial “WikiComet.” The interviews and field notes from this scenario serve as the basis for a thicker analytic account of how protocol rules, mediating practices, and second-order effects interact over time.

\subsubsection{Blue Build One (2025–2125): Defining a 100-Year Preservation Strategy}

The table began by selecting Wikipedia as the knowledge artifact to preserve, a choice that stemmed from a shared sense that Wikipedia approximates a “memory of civilization”: a living, collaboratively produced knowledge base that is both mundane and infrastructurally central to contemporary sense-making. Participants agreed that preservation should focus on knowledge that has ongoing use or interpretability. As one participant noted in the interview, knowledge that no longer connects to any plausible future use “belongs in a museum,” not in an active, resource-intensive preservation system. This early distinction between “living” and “museum” knowledge became an important design principle for prioritizing what should be carried forward.

The team’s 100-year preservation strategy combined technological and social components. On the technical side, participants proposed: atom-level etching of the full Wikipedia corpus on durable substrates; large-format, human-readable instructions in multiple languages; redundant copies distributed across jurisdictions, cold storage, and off-world locations;  and an artificial “WikiComet” that would carry a canonical copy and return on a 100-year orbit. On the social side, they imagined a modular, consensus-driven infrastructure, rooted in overlapping communities: open-source contributors, libraries, universities, and blockchain-based stewards. In the interview, participants described this as a “diffused ownership” model: no single institution controls the archive; instead, a network of actors maintains it based on shared protocols.

From a Protocol Futuring perspective, this phase reveals the first-order rules of the design: what qualifies as worth preserving (relevance and interpretability); how authenticity is defined (canonical copies, clear instructions, cryptographic assurances); and how governance is structured (consensus-based, modular, decentralized). The relay structure then shows how these rules are interpreted and transformed by later teams.

\subsubsection{Red Crisis One (2120–2125): Authenticity Under Siege}

The next team continued the story arc by introducing a plausible crisis that would disrupt the Blue Build 1 preservation effort. The team reinterpreted the design around a neo-Heideggerian movement. This movement argued that a single canonical encyclopedia represented a dangerous narrowing of the possible world—“one truth” as a form of existential extinction. Their intervention democratized micro-engraving tools and encouraged mass production of atom-level Wikipedia forgeries. A new social norm emerged: any claim to authenticity should be met with competing claims.

By the end of this five-year arc, the original strategy to replicate Wikipedia widely to secure authenticity had inverted. Redundancy became the substrate for epistemic jam. In interview reflections, participants described this as a core design lesson: a protocol built solely for truth-preservation, without anticipating adversarial reinterpretation, can be reappropriated into a political struggle in which authenticity itself collapses into contestation.

\subsubsection{Blue Build Two and Red Crisis Two (2125–2425): Hash Wars and Environmental Entanglement}

Subsequent teams attempted to repair the jam by introducing new layers of protocol: comet-based “carbon dating” of engraved replicas, sandboxed LLMs to reconstruct pre-LLM word distributions, and a blockchain-verified hash of the “oldest known Wikipedia.” Interviewed participants described this as an attempt to build a “protocol hardness”---a durability that seeks to anchor truth in mechanisms that could not easily be spoofed or reinterpreted. At the same time, the social milieu shifted. AI researchers, amateur astronomy clubs, and archivists formed a coalition around maintaining a “clean” pre-LLM dataset. This coalition reinforced the idea that some slices of knowledge must remain stable referents, even as everything else is up for reinterpretation.

The next crisis moved the story to a planetary scale. A volcanic event released centuries of e-waste into the atmosphere, creating an electromagnetic “e-distortion stratosphere” that scrambled radio communication between space and Earth. Hash comparison became energy-intensive and politically contentious. Malicious actors flooded the network with adversarial hashes; legislation restricted large-scale computation due to environmental impact.

By 2425, distorted information had become culturally normalized, while undistorted data was derided as naïve or heretical. In this stage, a protocol designed to verify authenticity through intensive hash comparison and space-based signaling became entangled with heavy compute demands, regulatory oversight, and adversarial exploitation, ultimately producing second-order effects such as environmental externalities, political clampdowns, and a cultural turn that celebrated distortion over fidelity. The interviews helped clarify this trajectory as a risk-aware framing of long-term preservation: participants explicitly discussed political, economic, social, legal, and environmental constraints as integral to protocol design. Crisis recovery was not just about restoring systems, but about maintaining legitimacy under shifting constraints.

\subsubsection{Blue Build Three and Red Crisis Three (2425–3025): Ritualization, Relevance, and Myth}

The last Blue Build team imagined a partial technical recovery: atmosphere-cleaning systems that gradually restored communication fidelity, energy-efficient hashing methods, and policy reversals that reopened the possibility of high-precision verification. Yet these technical gains were paired with a cultural and regulatory backlash. Following a major public-policy failure attributed to distorted data, “distortionist” practices were increasingly framed as threats to collective decision-making, prompting legal restrictions and renewed emphasis on data integrity. Preservation infrastructures had become intertwined with public accountability regimes and shifting moral expectations around truthfulness.

The final crisis layer reintroduced uncertainty rather than closure. A miscalculation in the long-term orbital trajectory of the preservation comet created widespread anxiety about the system’s safety and governance. Responsibility for the error was unevenly assigned, and archivists became symbolic targets of frustration over perceived technocratic overreach. At the same time, the scenario depicted communities forming new interpretive traditions around surviving fragments of the preservation system—treating certain artifacts or data snapshots as touchstones for shared identity and meaning. Rather than descending into spectacle, this represented a familiar sociotechnical pattern: when infrastructures fail or become opaque, preservation protocols can be reabsorbed into cultural practices, generating new forms of ritualization, symbolic value, or mythologized histories. What began as a technical design for knowledge preservation ended as cultural infrastructure—an object of devotion, blame, and collective imagination.

\subsection{Design Principles and Mediating Practices}
Interpretation occurred at multiple moments across the relay. During the workshop, participants themselves interpreted and reworked inherited protocol fragments under constrained conditions. After the workshop, the research team conducted a second-order analytic pass, drawing on field notes, the evolving documents, and follow-up interviews to trace how protocol rules, mediating practices, and second-order dynamics emerged across rounds. Minimal validity checks included adversarial critique built into the Red Crisis phases, participant reflection during the final debrief, and post-hoc elicitation interviews that allowed participants to confirm, contest, or reinterpret the researchers’ analytic readings.

The authors' participant-observer role, supported by field note reflection and follow-up interviews, enabled us to surface how participants made sense of their own design decisions as they enacted them. First, participants repeatedly emphasized relevance and interpretability as key criteria for long-term preservation. Interviewee 1 argued that not all knowledge should remain infrastructurally “live”; some artifacts should appropriately become “museum pieces.” This distinction sharpened our understanding of Protocol Futuring as more than an exercise in imagining static artifacts, but anticipating how knowledge will change in meaning, value, and usability over time, and to design protocols that can accommodate these shifts without collapsing into either rigid fidelity or indiscriminate accumulation.

Second, participants described infrastructure as modular, consensus-based, and characterized by diffused ownership across communities. This resonates with decentralized governance traditions, yet the relay format introduced a distinctive requirement: teams had to inherit and work with partially formed protocols rather than design from scratch. Doing so required mediation, negotiation, inference, and simplification. These mediating practices—rather than the initial technical choices—became the loci where drift accumulated and second-order effects emerged.

Third, participants treated crisis and recovery as intrinsic to protocol design. Several emphasized the need for “recovery protocols” capable of restoring function after disruption without sacrificing interpretability. This anticipatory stance toward failure differentiates Protocol Futuring from more static forms of design fiction, which often portray artifacts as working as intended within a closed narrative world. Here, resilience, repair, and flexible adaptation were core design considerations.

Fourth, participants explicitly noted how philosophical and cultural attitudes—such as skepticism toward historical truth or tendencies toward artistic reinterpretation—might shape the long-term fate of preservation systems. Rather than assuming fidelity as a universal good, they saw value in designing infrastructures that honor historical authenticity while permitting creative reimagination. This orientation positions Protocol Futuring as a way to explore how future publics might negotiate the tension between accuracy and meaning.

Finally, challenges around collaboration and communication surfaced repeatedly through metaphors of “different islands.” Because teams worked asynchronously through textual handovers, misinterpretation and semantic drift were routine features of the exercise. Participants proposed concepts such as “lingua defense”—shared languages, standards, and interpretive conventions—as cultural protocols to mitigate these breakdowns. This insight reframes communication and consensus mechanisms as preservation infrastructures in their own right, revealing how much interpretive labor underpins even seemingly technical systems.

\section{Discussion}
We now reflect on what Protocol Futuring offers for understanding sociotechnical infrastructures and for methodological practice. We focus on three themes: (1) temporality and second-order dynamics; (2) the relation between protocol futuring, design futuring, and infrastructural futuring; and (3) methodological implications, limitations, and future directions.

\subsection{Temporality, Second-Order Dynamics, and Protocol “Memory”}

\added{
In Knowledge Futurama, participants collectively designed archival protocols intended to persist for centuries. The resulting fictions centered not only on immediate trade-offs (e.g., privacy vs. access, centralization vs. federation) but also on how those choices sediment into long-term trajectories: protocols accumulating layers of patches, losing legibility, or being repurposed for unexpected ends. This echoes \citet{Ribes2007Tensions}’s long-term planning of infrastructures, where tensions between flexibility and stability, project timescales and infrastructural timescales, become central. Our speculative archive worlds dramatized, for instance, how a global archive protocol might gradually drift away from its founding rationale as institutional memory erodes and new stakeholders reinterpret its rules. The second-order dynamics surfaced in Knowledge Futurama should therefore be read as reflective hypotheses—patterns made thinkable through speculative practice and paired with candidate observables for future empirical follow-up, rather than as determinate claims about how archival infrastructures will evolve.
}

Seen through STS work on temporality, these stories can be understood as miniature sociotechnical imaginaries attached to particular protocol regimes. \citet{jasanoff2015dreamscapes} argue that large-scale projects—from nuclear power to smart grids—are guided by collectively held imaginaries of national futures. In our method, participants effectively articulate competing imaginaries of protocolized futures: worlds where archival AIs act as guardians of planetary memory versus worlds where protocol-enforced scarcity creates exclusion and information black markets. Protocol Futuring thus surfaces how different imaginaries are encoded into protocol choices, and how those imaginaries, in turn, shape what second-order phenomena are even thinkable.

Our scenarios also foreground protocol “memory”: the ways in which protocols remember (or fail to remember) the world that produced them. Bowker’s work on biodiversity databases shows how infrastructures embody particular classificatory decisions that persist long after their original context has faded \cite{bowker2000biodiversity}. Similarly, the archive stories exhibit situations where only a small priesthood remembers why certain safeguards exist by the year 2500. Participants proposed design responses—such as embedding narrative origin stories or mythologies into code comments, protocol documents, or ritual practices—that explicitly tie technical rules to their founding controversies. This points toward protocol design strategies that treat documentation, storytelling, and education as first-class components of long-term governance, not mere afterthoughts.

Finally, Protocol Futuring helps articulate potential temporal failure modes of protocols as interpretive hypotheses for further inquiry. Through our fictions, we can distinguish candidate patterns such as: (a) drift, where practice may diverge from formal rules; (b) jam, where incompatible incentives or overloaded dependencies could lead to contention and deadlock; and (c) ossification, where the cost or risk of change may become so high that even beneficial modifications are repeatedly deferred. These second-order dynamics are not predictions but interpretive hypotheses generated through speculative practice—plausible trajectories that warrant empirical investigation. Each pattern maps onto potentially observable phenomena in real infrastructures (e.g., workarounds and shadow systems, prolonged standardization disputes, or entrenched legacy protocols). Future empirical work can focus on candidate observables such as divergence between protocol text and workaround artifacts (drift), frequency and distribution of change proposals (ossification), or patterns of stakeholder contestation (jam).

\subsection{From Design Futuring and Infrastructural Futuring to Protocol Futuring}

 Protocol Futuring is not a substitute for infrastructural inversion, protocol ethnography, or policy prototyping; it complements these approaches by enabling inquiry under conditions where infrastructures are emergent, opaque, or not yet materially instantiated. Infrastructural inversion \cite{bowker1994science} reveals the hidden labor and classificatory schemes underpinning existing systems—but it presupposes a working infrastructure to invert. Protocol ethnography similarly depends on observable practices. Policy prototyping surfaces governance consequences, but is typically constrained to near-future institutional imaginaries. In contrast, Protocol Futuring is most valuable when protocols are nascent, rapidly evolving, or too abstract to have accumulated empirical traces. It provides a systematic way to explore long-range second-order effects when no real-world data yet exists, and to make consequential drift, contestation, or ossification experientially accessible to participants.

\added{A concrete example from Knowledge Futurama illustrates this complementarity. }
If one were to study Wikipedia's current archival practices through infrastructural inversion or protocol ethnography, one might surface the labor of verification, disputes over authority, governance bottlenecks, funding dependencies, and tensions between openness and control. These approaches make visible the hidden work and politics of existing archival infrastructures. What Protocol Futuring uniquely offers is that it extends this analysis by making visible how these same protocol rules transform over centuries of reinterpretation. Through speculative relay and crisis inheritance, participants surfaced dynamics that cannot yet be empirically observed. Rather than revealing what archives are, Protocol Futuring reveals what archive protocols become as they traverse long temporal horizons, shifting social imaginaries, and cumulative reinterpretation. 

Design futuring and critical/speculative design have historically emphasized form-giving: using material artifacts—objects, interfaces, films—to tell stories about possible futures \cite{dunne2013speculative, fry2009design}. \citet{Sanders2010framework} highlight how materials and design things become shared platforms for participatory exploration and ongoing infrastructuring. In this tradition, speculative artifacts can generate knowledge by embodying tensions, inviting interpretation, and revealing otherwise latent assumptions.

Protocol Futuring diverges from this artifact-centric focus by treating rules, standards, and coordination mechanisms as the primary design material. Instead of asking “What if this object existed?”, we ask “What if this protocol governed interactions?” Yet this does not mean form and material disappear. In our practice, protocols are materialized through a variety of representational forms: pseudo-RFCs, diagrams, decks of rule cards, AI-generated films, and workshop enactments. These are not incidental; they are the forms through which participants grasp, negotiate, and contest protocols. Future work can deepen this connection by explicitly drawing on material speculation—e.g., building physical or mixed-reality artifacts whose behavior is driven by alternative protocol logics—to explore what is gained or lost when rules are embodied in different media.

Infrastructural futuring and infrastructural speculations, by contrast, deliberately place lifeworlds, infrastructures, and repair practices at the center of speculative concern \cite{wong2020infrastructural,Harvey2012Topological}. These approaches leverage infrastructural inversion to reveal how dependencies, maintenance work, and governance arrangements shape everyday experience. Our method aligns with this orientation but operates at a more meso-level: the scale of protocol regimes that sit between individual artifacts and entire infrastructures. Where infrastructural futuring might imagine future energy grids or mobility systems, Protocol Futuring drills into specific protocol levers within those systems—archival standards, consensus rules, access control schemes—and asks how changes at that level reverberate through infrastructures and lifeworlds over time.

This suggests a decision rule for method choice:
\begin{itemize}
    \item When infrastructures are already instantiated and partially legible, infrastructural inversion and ethnographic study may be the most appropriate first step.
    \item  When infrastructures are emergent, highly abstracted, or institutionally opaque, or when we wish to explore speculative regimes that cannot yet be observed empirically, Protocol Futuring can extend infrastructural thinking into the prospective long-now of protocols.
\end{itemize}
In this sense, Protocol Futuring does not replace infrastructural futuring but adds a rule-centric lens that can be combined with it—for example, by pairing speculative protocol scenarios with later ethnographic or policy-prototype work around actual standards processes.

\subsection{Limitations and Scope}
Protocol Futuring, like other speculative and research-through-design approaches, operates through reflective enactment rather than empirical verification. Its limitations therefore lie not in a lack of actionability or representativeness, but in its situated interpretive scope. First, PF foregrounds long-term second-order dynamics—drift, jam, ossification—which means its insights are most applicable to infrastructures and governance systems characterized by uncertainty, plurality, and extended temporal horizons. It is less suited for tightly bounded systems where short-cycle empirical data or formal verification methods are more appropriate.

Second, as with all speculative methods, Protocol Futuring reflects the positionalities and imaginations of its participants. Our workshops tended to attract people already comfortable with technology and futures thinking; perspectives from those marginalized by existing infrastructures or skeptical of future-oriented design were less present. Because speculative infrastructures risk legitimizing the imaginaries they articulate, facilitators must engage critically with whose futures are being made possible and whose remain unaddressed. Reflexive framing, intentional outreach, and co-design with affected communities are necessary to mitigate these exclusions.

Third, Protocol Futuring captures relational and processual patterns rather than producing prescriptive solutions. Its contributions are generative—clarifying tensions, surfacing potential failure modes, and expanding the design space for protocol governance—rather than determinative. This generativity is a strength, but it also means PF should be complemented with empirical approaches such as infrastructural ethnography, protocol analysis, simulation, or field deployment when more concrete design commitments are required.

Finally, future research should extend PF into longitudinal or hybrid formats. One promising direction is integrating speculative protocol worlds with empirical studies of existing infrastructures, or iteratively testing protocol prototypes within real governance bodies. Such combinations may strengthen both the analytic reach and practical utility of PF by anchoring long-term speculative insights in real sociotechnical constraints.

\section{Conclusion}

We set out to explore how speculative design can engage the often invisible yet critical world of protocols—the rules and infrastructures that underpin sociotechnical systems. To this end, we introduced Protocol Futuring as a methodological contribution that blends critical design, participatory foresight, and infrastructure studies. For HCI, and especially the Critical Computing community, Protocol Futuring provides a way to examine issues of power, equity, and sustainability before technologies are fully realized. It complements traditional user-centered design by surfacing visionary use cases and potential misuses, and it serves an educational and public engagement role by communicating complex research questions in accessible form. We see opportunities to integrate Protocol Futuring into participatory design (for example, as part of a smart-city project to elicit citizen concerns) or speculative prototyping (for instance, building a low-fidelity prototype of a fictional protocol to test a specific dynamic). The temporal flexibility of fiction—its ability to fast-forward and rewind—offers a unique advantage for infrastructures that cannot be readily tested in a lab because of their scale or long time horizons.

\begin{acks}
This work was supported by the Summer of Protocols program, funded by the Ethereum Foundation. We thank Venkatesh Rao, Tim Beiko, and Timber Stinson-Schroff for their guidance and support throughout the program.

\paragraph{Disclosure of the usage of LLM}
We used ChatGPT 5 to facilitate the writing of this manuscript. The usage includes:
\begin{itemize}
    \item Summarizing literature;
    \item Polishing existing writing;
    \item Proofreading for grammar and spelling.
\end{itemize}

\end{acks}

\balance

\bibliographystyle{ACM-Reference-Format}
\bibliography{reference}

\appendix

\clearpage 
\section{Knowledge Futurama Scenario Worksheet}

\subsection*{BLUE BUILD ONE (2025--2125)}

\textbf{Team members:}  \hrulefill

\noindent\textbf{What is the KNOWLEDGE ARTIFACT you want to preserve for 1000 years?} \emph{(e.g., genomic data, Bible, Wikipedia\ldots)}

\hrulefill

\noindent\textbf{What is the TECHNOLOGY STACK for the preservation?} \emph{(e.g., IPFS, crystal launched into space, paper copies in vaults, trained LLM on hardened computer\ldots)} Describe as a bullet list (7 items max).
\begin{itemize}
  \item \dots
  \item \dots
  \item \dots
  \item \dots
  \item \dots
  \item \dots
  \item \dots
\end{itemize}

\noindent\textbf{What is the SOCIAL MILIEU for the preservation?} \emph{(e.g., secret cult of priests, open-source community, universities)} Describe as a bullet list (7 items max).
\begin{itemize}
  \item \dots
  \item \dots
  \item \dots
  \item \dots
  \item \dots
  \item \dots
  \item \dots
\end{itemize}

\noindent\emph{At this point hand over the document to the next team!}

\subsection*{RED CRISIS ONE (2120--2125)}

\noindent\textbf{Team members:} \hrulefill

\noindent\textbf{Describe the crisis affecting BLUE BUILD ONE in this 5-year period.} Keep it sociologically realistic and technologically plausible. The crisis should threaten and reshape BLUE BUILD ONE in unexpected ways, but not destroy it ($<$10 bullets).
\begin{itemize}
  \item \dots
  \item \dots
  \item \dots
  \item \dots
  \item \dots
  \item \dots
  \item \dots
  \item \dots
  \item \dots
  \item \dots
\end{itemize}

\noindent\textbf{What is the state of BLUE BUILD ONE at the end of the crisis in 2125?} Describe as circumstances/state conditions (7 bullets max).
\begin{itemize}
  \item \dots
  \item \dots
  \item \dots
  \item \dots
  \item \dots
  \item \dots
  \item \dots
\end{itemize}

\noindent\emph{At this point hand over the document to the next team!}

\subsection*{BLUE BUILD TWO (2125--2425)}

\noindent\textbf{Team members:} \hrulefill

\noindent\textbf{What is the TECHNOLOGY STACK for the preservation?} This can get more science fictional than BLUE BUILD ONE, but do not break the laws of physics. Describe as a bullet list (7 items max).
\begin{itemize}
  \item \dots
  \item \dots
  \item \dots
  \item \dots
  \item \dots
  \item \dots
  \item \dots
\end{itemize}

\noindent\textbf{What is the SOCIAL MILIEU for the preservation?} Factor in BLUE BUILD ONE and RED CRISIS ONE. Can be slightly transhumanist. Describe as a bullet list (7 items max).
\begin{itemize}
  \item \dots
  \item \dots
  \item \dots
  \item \dots
  \item \dots
  \item \dots
  \item \dots
\end{itemize}

\noindent\emph{At this point hand over the document to the next team!}

\subsection*{RED CRISIS TWO (2420--2425)}

\noindent\textbf{Team members:} \hrulefill

\noindent\textbf{Describe the crisis affecting BLUE BUILD TWO in this 5-year period.} Keep it sociologically realistic and technologically plausible. The crisis should threaten and reshape BLUE BUILD TWO in unexpected ways, but not destroy it ($<$10 bullets).
\begin{itemize}
  \item \dots
  \item \dots
  \item \dots
  \item \dots
  \item \dots
  \item \dots
  \item \dots
  \item \dots
  \item \dots
  \item \dots
\end{itemize}

\noindent\textbf{What is the state of BLUE BUILD TWO at the end of the crisis in 2425?} Describe as circumstances/state conditions (7 bullets max).
\begin{itemize}
  \item \dots
  \item \dots
  \item \dots
  \item \dots
  \item \dots
  \item \dots
  \item \dots
\end{itemize}

\noindent\emph{At this point hand over the document to the next team!}

\subsection*{BLUE BUILD THREE (2425--3025)}

\noindent\textbf{Team members:} \hrulefill

\noindent\textbf{What is the TECHNOLOGY STACK for the preservation?} This can get even more science fictional than BLUE BUILD TWO, but do not break the laws of physics. Describe as a bullet list (7 items max).
\begin{itemize}
  \item \dots
  \item \dots
  \item \dots
  \item \dots
  \item \dots
  \item \dots
  \item \dots
\end{itemize}

\noindent\textbf{What is the SOCIAL MILIEU for the preservation?} Factor in BLUE BUILD TWO and RED CRISIS TWO. Can be more transhumanist. Describe as a bullet list (7 items max).
\begin{itemize}
  \item \dots
  \item \dots
  \item \dots
  \item \dots
  \item \dots
  \item \dots
  \item \dots
\end{itemize}

\noindent\emph{At this point hand over the document to the next team!}

\subsection*{RED CRISIS THREE (3020--3025)}

\noindent\textbf{Team members:} \hrulefill

\noindent\textbf{Describe the crisis affecting BLUE BUILD THREE in this 5-year period.} Keep it sociologically realistic and technologically plausible. The crisis should threaten and reshape BLUE BUILD THREE in unexpected ways, but not destroy it ($<$10 bullets).
\begin{itemize}
  \item \dots
  \item \dots
  \item \dots
  \item \dots
  \item \dots
  \item \dots
  \item \dots
  \item \dots
  \item \dots
  \item \dots
\end{itemize}

\noindent\textbf{What is the state of BLUE BUILD THREE at the end of the crisis in 3025?} Describe as circumstances/state conditions (7 bullets max).
\begin{itemize}
  \item \dots
  \item \dots
  \item \dots
  \item \dots
  \item \dots
  \item \dots
  \item \dots
\end{itemize}

\noindent\textbf{Give this scenario a NAME:} \hrulefill

\noindent\emph{The team holding this document at this stage will narrate the whole story in the shareback in 3--5 minutes.}
\clearpage 

\section{Additional Case Studies}

\subsection{South Beast Asia}

\textbf{Context \& Goal}: South Beast Asia is a fictional world developed through a five-day ``Protocol Futuring'' workshop co-hosted by Seapunk Studios (a collective exploring solarpunk and decentralized futures in Southeast Asia) in April 2025. The workshop's title ``Khlongs \& Subaks'' hints at its setting: khlongs (canals in Thailand) and subaks (community-managed irrigation systems in Bali)---both are traditional infrastructures. The goal was to envision a future Southeast Asia (SEA) where these local infrastructure concepts meet advanced AI and networking protocols, under the theme of ``a fictional world haunted by AIs in all sorts of (mostly) friendly living embodied forms''. Essentially, it was an attempt to create a regionally grounded protocol design fiction, incorporating local cultures, ecologies, and socio-political contexts often underrepresented in mainstream tech futures. The involvement of Seapunk indicates a thematic leaning toward sustainable, community-oriented futures (solarpunk ethos) mixed with playful, creative elements (the name suggests a beastly, fantastical twist).

\textbf{Process \& Outputs}: A few dozen researchers and creators (from the region and beyond) participated. They engaged in collaborative worldbuilding, which included:
\begin{itemize}
    \item Defining key premise: AIs exist that are embedded in the environment (in animals, trees, rivers, as well as devices), giving rise to a world that feels alive with digital spirits---very much an overlap with Ghosts in Machines theme, but specifically localized.
    \item Determining rules/protocols of the world: For instance, one rule might be that all AI agents follow a ``hospitality protocol'' influenced by local customs---meaning any AI entering a new village must announce itself and respect certain boundaries (like spiritual guardians in folklore). Another might be a resource-sharing protocol among AIs to coordinate energy usage, echoing how subak systems share water equitably for rice farming.
    \item Creating characters and story arcs: They personified some AIs as characters (like an AI in the form of a water buffalo that manages rice field irrigation, or a shrine guardian AI that curates local history) and also human characters who interact with them (farmers, monks, children, hackers).
    \item Prototyping artifacts: The workshop produced thirteen short video vignettes (AI-generated or animated) that depict scenes in this world, a trailer summarizing the world, an expository essay describing its concepts, plus numerous concept images and fictional artifacts. This output was comprehensive enough to be considered a ``season'' of content for Seapunk's creative storyline releases.
\end{itemize}

\textbf{Themes \& Second-Order Insights}:
\begin{itemize}
    \item \textbf{Hybridization of Tradition and Technology}: South Beast Asia highlights how future protocols might co-evolve with existing cultural frameworks rather than overwrite them. Instead of erasing tradition, these design fictions interweave technical rules with long-standing practices—for example, a speculative hospitality protocol for AIs that updates the region’s customs of greeting and mutual respect. The second-order implication is that technology could be more peacefully adopted and sustainable when it adapts to local mores. Yet this raises further questions: who has the authority to define such norms, and how might they be enforced? If an AI fails to perform the required ritual, would it be punished or even banned from the network, echoing social forms of ostracism? This scenario reframes network governance as cultural governance applied to nonhuman agents, suggesting that the social sanctions that maintain human communities might one day regulate artificial ones as well.
    
    
    \item \textbf{Decentralization \& Resilience}: Because SEA has many islands and diverse communities, a key protocol design in the world was one of federation rather than top-down control. No single AI rules; many small ones coordinate. They used a protocol analogous to gossip protocols in distributed systems, or maybe something like the ActivityPub of AI agents. This architecture implies resilience (no single point of failure), but the fiction also looked at challenges: e.g., how do you resolve conflicts between local AI enclaves? One story involved a disagreement between a river AI and a land AI over diverting water for a town vs. for rice paddies during drought---essentially a local resource conflict that in human terms would be resolved by community negotiation. They had the AIs actually convene a mixed council with humans (enabled by some translation protocol) to mirror how conflicts are resolved in village meetings. The second-order effect shown is that protocols for AI-human collaboration become political: people have to create new governance bodies that include non-humans as stakeholders.
    
    \item \textbf{Climate and Ecology Integration}: South Beast Asia's ``beasts'' are often nature integrated---e.g., AI-enhanced animals or AI spirits of forests. A major narrative undercurrent is climate adaptation; AI protocols are depicted as helping regrow mangroves, manage coral reefs, or anticipate typhoons, but also the risk that if they malfunction or follow misguided objectives, they could cause environmental harm at scale. One speculative element was climate justice protocols---for example, an AI that enforces carbon reduction by modulating energy access. If an entity is over their carbon budget, their devices politely dim or throttle (the AI ghost of climate literally haunting energy grids). This raises social acceptance issues: the story might show some people taming or tricking the ghost (e.g., painting a sensor black to hide emissions), a cat-and-mouse dynamic. The second-order concern: protocols embodying environmental values might need coercive power that not everyone consents to, leading to tensions in freedom vs. survival trade-offs.
\end{itemize}

\end{document}